**Forecasting the value of battery electric vehicles compared to internal combustion engine vehicles: the influence of driving range and battery technology**

JongRoul Woo[a,*] and Christopher L. Magee[a,b]

[a] *Institute for Data, Systems, and Society, Massachusetts Institute of Technology, 77 Massachusetts Avenue, Cambridge, MA 02139-4307, United States*

[b] *SUTD-MIT International Design Center, Massachusetts Institute of Technology, 77 Massachusetts Avenue, Cambridge, MA 02139-4307, United States*

[*] Corresponding author. E-mail: jroul86@mit.edu; Tel.: +1-617-386-3392

**Abstract**

Battery electric vehicles (BEVs) are now clearly a promising candidate in addressing the environmental problems associated with conventional internal combustion engine vehicles (ICEVs). Accordingly, governments in many countries have promoted the consumer adoption of BEVs by providing financial incentives and automobile manufacturers are accelerating their efforts to develop BEVs. However, BEVs, unlike ICEVs, are still not widely accepted in the automobile market but continuing technological change could overcome this barrier. The aim of this study is to assess and forecast whether and when design changes and technological improvements related to major challenges in driving range and battery cost will make the user value of BEVs greater than the user value of ICEVs. Specifically, we estimate the relative user value of BEVs and ICEVs resulting after design modifications to achieve different driving ranges by considering the engineering trade-offs based on a vehicle simulation. Then, we analyze when the relative user value of BEVs is expected to exceed ICEVs as the energy density and cost of batteries improve because of ongoing technological change. Our analysis demonstrates that the relative value of BEVs is lower than that of ICEVs because BEVs have high battery cost and high cost of time spent recharging despite high torque, high fuel efficiency, and low fuel cost. Moreover, we found the relative value differences between BEVs and ICEVs are found to be less in high performance large cars than in low performance compact cars because BEVs can achieve high acceleration performance more easily than ICEVs. In addition, this study predicts that in approximately 2050, high performance large BEVs could have higher relative value than high performance large ICEVs because of technological improvements in batteries; however low performance compact BEVs are still very likely to have significantly lower user value than comparable ICEVs until well beyond 2050.

Keywords: Electric vehicle; Internal combustion engine vehicle; Battery technology; Driving range

# 1. Introduction

Battery electric vehicles (BEVs) are energy efficient and produce zero tail pipe emissions as they use electric motors and motor controllers instead of internal combustion engines for propulsion and battery packs instead of fuel tanks to store energy. Because of these technical characteristics, BEVs have received considerable attention as the solution to address the environmental problems such as climate change and air pollution associated with conventional internal combustion engine vehicles (ICEVs). Accordingly, governments in many countries have promoted the consumer adoption of BEVs by providing both financial and non-financial incentives such as tax exemptions, expansion of related infrastructure, free parking, discounted/free toll, and use of high-occupancy-vehicle lanes [1-5]. Moreover, automobile manufacturers are speeding up their efforts to develop BEVs [6].

However, BEVs, unlike ICEVs, are not yet widely accepted in the automobile market. Moreover, the acceptance appears more advanced in higher performance high priced vehicles than it is in the mass market. Specifically, in 2017, 1.2% of new cars sold in the U.S. were electric vehicles but 24.1% of the electric vehicles sold in U.S. were Tesla Model S and X which start at $74,500 (only about 1.5% of ICEVs sold in U.S. were priced at more than $74,500) [7]. Many researchers have noted that key barriers to BEV market penetration are high battery costs and limited driving ranges [8-14]. However, recent technological developments in BEVs provide an encouraging signal for both areas [15, 16].

A working assumption in this paper is that BEVs will be widely accepted in the automobile market only when the value or utility which BEVs can give to consumers is as high or higher than ICEVs[1]. Thus, in order to assess the future of BEVs in the market, it is important to investigate how design changes and technological improvements related to driving range and batteries in BEVs affect the user value of BEVs and whether these effects are likely to make the user value of BEVs greater than that of ICEVs.

ICEVs and BEVs are essentially mobile machines that transport people or cargo. However, the user value of ICEVs and BEVs are difficult to evaluate comprehensively because consumers generally value an extensive set of both objective and subjective attributes. In this study, we develop an index to evaluate the *relative* value of ICEVs and BEVs based on the vehicle attributes which are distinguishable because of the fundamental technological differences in ICEVs and BEVs.

Previous studies have analyzed the marginal effects of vehicle attributes such as price and driving range on the user value of ICEVs and BEVs by modeling consumer choice behavior and demand for the automobile market among existing products [17-24]. However, it is problematic to assess how design changes and technological improvements in the mid- and long-term affect the user value of ICEVs and BEVs if one only considers demand side empirical models. Thus, we go beyond demand considerations and treat engineering trade-offs in ICEVs and BEVs using a relatively simple user value index.

---

[1]Carbon pricing or other incentives to deal with climate change will change the user value of BEVs if they are enacted and if electricity production systems become more carbon-free than at present.

The technical differences between ICEVs and BEVs affect consumer attributes, trade-offs among these attributes, and the relative user value of ICEVs and BEVs. Therefore, this study estimates changes in the vehicle attributes and relative user value of BEVs and ICEVs resulting from design changes to achieve driving ranges not currently available. This is done while considering the engineering trade-offs based on a vehicle simulation. Then, we analyze when the relative user value of BEVs is expected to exceed ICEVs as the energy density and cost of batteries improve because of ongoing technological change. We ignore much other technological change which more or less equally affects ICEVs and BEVs.

The remainder of this paper is organized as follows. Section 2 describes our relative user value index for ICEVs and BEVs and the simulation methods we use in this study. Section 3 presents the simulation and sensitivity test results. Section 4 interprets the results and discusses their implications. Section 5 provides conclusions.

## 2. Methods
### 2.1. Relative user value index of ICEVs and BEVs

In this study, we develop an index to evaluate the relative user value of ICEVs and BEVs. There are numerous vehicle attributes that provide utility to the user of the vehicle. However, this study develops the relative user value index based on vehicle attributes that characterize differences between ICEVs and BEVs in user value according to the key technical differences. The fundamental technical difference between ICEVs and BEVs is that the core technological domains are different in the energy storage and energy conversion of the vehicle. ICEVs store energy by storing gasoline in fuel tanks, while BEVs use batteries to store electrical energy in the form of chemical energy; ICEVs use internal combustion engines and BEVs use electric motors and motor controllers to gain propulsion to transport people or cargo based on the stored energy. These fundamental differences in the technology used in ICEVs and BEVs affect some consumer attributes and thus the relative user value of ICEVs and BEVs. Specifically, because current technology-level batteries are expensive but inferior to gasoline in terms of energy density, BEVs tend to be more expensive, have shorter driving range, and have smaller internal spaces than comparable

ICEVs. Moreover, BEVs take a longer time to charge the batteries compared to fueling ICEVs with gasoline. These factors reduce the relative user value of BEVs versus ICEVs. On the other hand, BEVs have an advantage over ICEVs in terms of accerleration performance and operating cost and these factors increase the relative user value of BEVs versus ICEVs because BEV electric motors have higher torque and are more efficient than internal combustion engines.

We chose the vehicle attributes and the appropriate metrics to construct the relative user value index based on such differences between ICEVs and BEVs for technology and attributes as given in Table 1. We ignore other vehicle attributes that are basically unchanged between ICEVs and BEVs.

**Table 1. Vehicle attributes and metrics for the relative user value index.**

| Consumer need | Related vehicle attribute | Metric |
| --- | --- | --- |
| Low cost | Initial vehicle purchase cost | Manufacturer's suggested retail price ($) |
| | Operating cost | Fuel cost ($) |
| High utility | Acceleration performance | 0 to 60 mph acceleration time $(s)^{-1}$ |
| | Driving range | Cost of time spent refueling or recharging ($) |
| | Full refueling or charging time | |
| | Passenger and cargo space | Interior volume (cubic feet) |

We construct the relative user value index for ICEVs and BEVs by combining the vehicle attributes and their metrics in Table 1 as equation (1)[2]. This linear approach effectively gives equal weight to each attribute; although more complex approaches are possible [26-28], there is no information from which to estimate the extra parameters called for so the linear equation is the best approximation. The relative user value index given in equation (1) technically represents the ratio of moving a spatial carrying capacity a unit distance with a given performance capability per unit of ownership cost. In this study, this index is calculated to measure the relative value of ICEVs and BEVs for a user driving 13,500 miles per year[3].

---

[2] An and DeCicco [25] developed a measure of vehicle value given by combining fuel economy with both performance and size and called it Performance-Size-Fuel economy Index (PSFI). Our relative user value index for ICEVs and BEVs can be viewed as an extension of PSFI.
[3] On average, annual miles per driver in the U.S. = 13,500 miles [29].

(1)
$$\text{Relative user value index}\left(\frac{ft^3 \cdot mile}{\$ \cdot s}\right) = \frac{\text{Interior volume}(ft^3) \times \text{Acceleration performance}(1/s)}{\text{Total cost of ownership}(\$)/\text{Miles traveled}(miles)}$$
$$= \frac{\text{Interior volume}(ft^3) \times \text{Acceleration time}(s)^{-1}}{\frac{\text{Initial vehicle cost}(\$) + \text{Fuel cost}(\$) + \text{Cost of time spent fueling or charging}(\$)}{\text{Miles traveled}(miles)}}$$

The toal cost of ownership in equation (1) is the cost of maintaining the vehicle for a year[4] and is calculated by adding initial vehicle cost, fuel cost, and cost of time spent refueling or recharging. Specifically, the initial vehicle cost is the vehicle depreciation cost for one year and is calculated by multiplying the manufacturer's suggested retail price (MSRP) ($) by the depreciation rate (%). In this study, the percentage rate of vehicle depreciation is assumed to be 20%[5]. The fuel cost is then calculated by multiplying fuel consumption (gallon/mile or kwh/mile), fuel price ($/gallon or $/kwh)[6], and vehicle miles driven for one year. The cost of time spent refueling or recharging represents the value loss due to refueling or recharging time [31]. The cost of time spent fueling for ICEVs and the cost of time spent charging for BEVs is calculated using equation (2) and equation (3), respectively.

(2)
$$\text{Cost of time spent fueling for ICEVs}(\$) = \frac{M}{R_1} \times W \times H_s$$

(3)
$$\text{Cost of time spent charging for BEVs}(\$) = \frac{M}{R_2} \times \{W \times H_h \times f + W \times H_s \times (1-f)\}$$

Here, $W$ represents the value of driver's time and it is assumed to be 28 $/hour [32]. $M$ represents vehicle miles driven for one year. $R_1$ and $R_2$ represent driving range on a fully charged or filled fuel. It is assumed that a driver uses 80% of full fuel before refueling or recharging. Then, $R_1$ is calculated as 80% × fuel tank capacity (gallon) / fuel consumption (gallon/mile) and $R_2$ is calculated as 80% ×

---

[4]The auto manufacturers generally insure a BEV battery life for 8 years and 100,000 miles. In particular, Tesla's warranty covers 8 years and unlimited miles on battery and drive train. Thus, this study does not include the cost associated with battery life when calculating the toal cost of ownership of BEVs.
[5]On average, a new car loses about 20% of its value in its first year of ownership in the U.S.
[6]In 2016, Average U.S. electricity price = 0.095 $/kwh and average U.S. gasoline price = 0.061 $/kwh [30].

battery capacity(kwh) / fuel consumption(kwh/mile). $f$ is the percentage of home and workplace charging and the baseline assumption is that BEV drivers do 80% of their charging at home and work[7]. $H_h$ is the time required per recharging at home. We assume that the home or workplace base recharging is done during slack periods, so $H_h$ is defined to be only connect/disconnect time, not the entire time required for recharging. Then, it is assumed that connecting and disconnecting require only 1 minute. $H_s$ is the time required per refueling or recharging at refueling or recharging stations and it is assumed that external refueling or recharging require 10 more additional minutes during refueling or recharging[8]. Table 2 shows examples of calculating the relative value index for selected actual ICEVs and BEVs (data from [35]).

Figure 1 shows the relationship between the relative value index and sales of vehicles sold in the U.S. from 2013 to 2016. The vehicle specficiations by model (interior volume, energy consumption, acceleration time, charging time, battery size) for calculating the relative value index of ICEVs and BEVs were collected from the fuel economy data of the U.S. Environmental Protection Agency [35] and the vehicle sales and MSRP by model in the U.S. are collected from the Ward's Automotive Yearbooks [36]. Figure 1 shows correlation - vehicles with higher index values tend to be sold more in the market supporting the index as developed. However, the relative value index does not fully explain vehicle sales because many attributes other than those attributes included in the relative value index play a role in consumer vehicle purchase decisions such as aesthetic value, safety, warranty, brand reputation, resale value, and availability of refuelling infrastructure [37]. In our further use of the index, we assume that such attributes are not affected by the BEV/ICEV decision.

---

[7]Current BEV drivers do more than 80% of their charging at home and work [33].
[8]Gasoline pump usually takes 1 minute to fuel 6 gallons of gasoline [34]. Tesla Model S with 85 kWh battery pack needs about 285 minutes to be 100% charged and Nissan Leaf with 25 kWh battery pack needs about 300 minutes to be 100% charged [35].

**Table 2. Examples of relative value index calculation**

|  |  | ICEVs | | BEVs | |
|---|---|---|---|---|---|
|  | Brand | Honda | Honda | Nissan | Nissan |
|  | Model | 2016 Civic | 2013 Civic | 2016 Leaf | 2013 Leaf |
| **Relative value index ($\frac{ft^3 \cdot mile}{\$ \cdot s}$) = A / B** | | **36.55** | **25.61** | **16.23** | **15.65** |
| **A. Interior volume / Acceleration time (ft$^3$/s) = (1) / (2)** | | **13.19** | **10.39** | **11.76** | **11.64** |
| (1) | Interior volume (ft$^3$) | 113.4 | 100.9 | 116.4 | 116.4 |
| (2) | Acceleration time (s) | 8.59 | 9.72 | 9.90 | 10.00 |
| **B. Total cost / miles traveled (2016$/mile) = C / 13500 miles** | | **0.3610** | **0.4055** | **0.7244** | **0.7437** |
| **C. First year total cost of new vehicle = D + E + F** | | **4873.63** | **5473.77** | **9778.78** | **10040.11** |
| **D. First year depreciation (2016$) = ((3) – (4)) × 20%** | | **3888.00** | **3906.95** | **4302.00** | **5010.13** |
| (3) | MSRP (2016$) | 19440 | 19535 | 29010 | 32776 |
| (4) | Federal tax credit (2016$) | - | - | 7500 | 7725 |
| **E. Total fuel cost (2016$) = (5) × (6) × 13500 miles** | | **779.25** | **1355.99** | **379.46** | **400.99** |
| (5) | Fuel price (2016$/kwh) | 0.0614 | 0.0992 | 0.0948 | 0.1024 |
| (6) | Energy consumption (kwh/mile) | 0.94 | 1.01 | 0.30 | 0.29 |
| **F. Cost of time spent refueling or recharging (2016$)** | | **206.37** | **210.83** | **5097.32** | **4629.00** |
| (7) | Fuel tank or battery capacity (kwh) | 417.93 | 444.89 | 25.00 | 21.75 |
| (8) | Energy consumption (kwh/mile) | 0.94 | 1.01 | 0.30 | 0.29 |
| (9) | Driving Range for full (miles) | 445 | 439 | 84 | 75 |
| (10) | Fueling or charging time for full (min) | 2.1 | 2.2 | 300.0 | 240.0 |

Note: all dollar amounts are adjusted to 2016 dollars by using the CPI (Consumer Price Index).

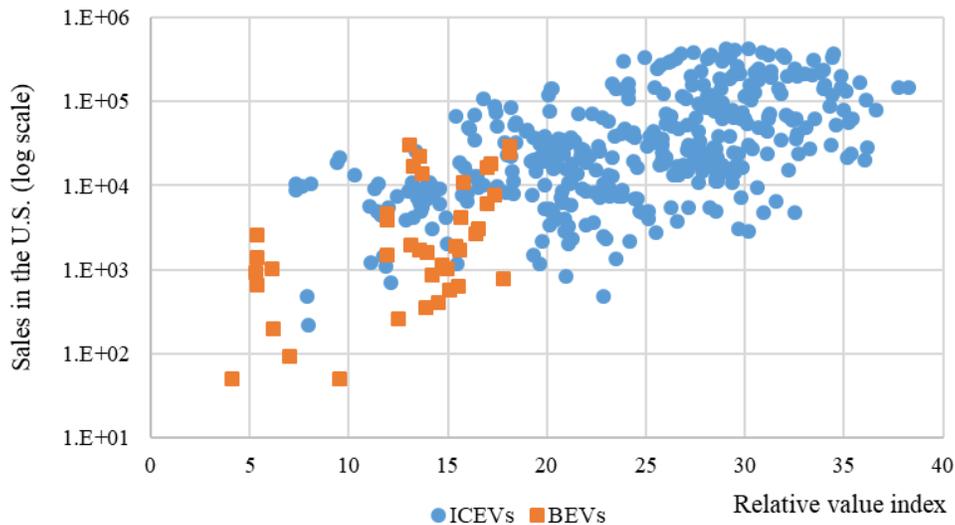

**Figure 1. Scatter plot between sales and relative value index of vehicles sold in the U.S. from 2013 to 2016.**

**2.2. Simulation of trade-offs in ICEVs and BEVs**

There are engineering trade-offs among the vehicle attributes which determine the relative value index of ICEVs and BEVs. In this study, we estimate changes in the vehicle attributes and relative user value index of BEVs and ICEVs accompanying changes of fuel tank size of ICEVs and battery size of BEVs by considering the engineering trade-offs in ICEVs and BEVs from a vehicle simulation based on NREL's FASTSim software[9]. Engineering trade-offs related to the change of fuel tank and battery size in BEVs and ICEVS are described in Figure 2. The change of fuel tank size of ICEVs or battery size of BEVs directly affects driving range, vehicle weight, interior size, and vehicle price, and then the resulting changes in vehicle weight indirectly affect acceleration performance and fuel consumption. Moreover, these direct and indirect effects modify the relative value index.

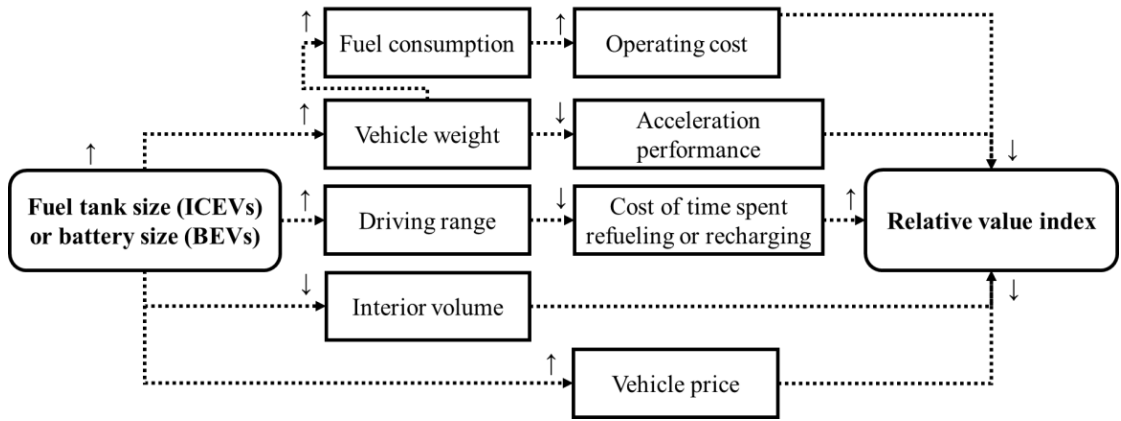

**Figure 2. Trade-offs in ICEVs and BEVs**

Among BEVs currently sold in the U.S. market, the Nissan Leaf (low performance compact car) and Tesla model S (high performance large car) are the most sold. We use these two vehicles as the starting point for two vehicle segments- a hypothetical low performance compact BEV which is a stand-in for mass market acceptance and a hypothetical high performance large BEV which is a stand-in for a high performance luxury BEV segment. Moreover, we conduct similar estimations/simulations based on a hypothetical low performance compact ICEV and a hypothetical high performance large ICEV which are comparable to the hypothetical BEVs, and then compare the results between BEVs and ICEVs in each of these two segments.

---

[9]We use NREL's Future Automotive Systems Technology Simulator (FASTSim) [38]. The FASTSim models ICEVs and BEVs and provides a simple way to compare powertrains and estimate the impact of technology improvements on vehicle efficiency, performance, and cost.

We select input parameters for hypothetical low performance compact and high performance large BEVs with battery specific energy and battery cost assumed to be at the 2016 level (Table 3). Then we also select input parameters for hypothetical low performance compact and high performance large ICEVs which are comparable to the low performance compact and high performance large BEVs (Appendix A).

**Table 3. Simulation input parameters for low performance compact and high performance large BEVs**

| Classification | Parameter | low performance compact BEV | high performance large BEV |
|---|---|---|---|
| Vehicle | Drag coefficient | 0.29 | 0.28 |
| | Frontal area (m$^2$) | 2.07 | 2.14 |
| | Vehicle glider mass (kg) | 768.77 | 809 |
| | Vehicle center of gravity height (m) | 0.53 | 0.53 |
| | Wheel base (m) | 2.59 | 2.82 |
| Motor | Motor power (kW) | 100 | 250 |
| | Motor peak efficiency | 0.90 | 0.93 |
| Battery | **Variable: Battery energy (kWh)** | **From 1 to 300** | **From 1 to 300** |
| | Battery specific energy (kWh/kg) | 0.25 | 0.25 |
| | Battery Cost (kWh/1000$) | 3.33 | 3.33 |
| | Markup factor* | 1.5 | 1.5 |
| Wheel | Number of wheels | 4 | 4 |
| | Tire radius (m) | 0.317 | 0.334 |

* In NREL's FASTSim software, the component costs are multiplied by a specified markup factor, which is set to 1.5 by default. This markup translates the cost to make the component to an estimated price impact on the vehicle [39].

## 2.3. Technological progress in battery technology

Although many researchers have pointed out that key barriers to BEV market penetration are high battery costs and limited driving ranges, battery technology is improving. Thus, in this study, we also simulate when the relative user value of BEVs could exceed ICEVs in the future due to ongoing technological changes in battery technology.

In this forecasting part of our study, we use a well-established approach usually referred to as the generalized Moore's Law (GML). In this formalism, technological performance is measured by output divided by price or other constraints such as weight [40]. In this study, we measure the technological performance of batteries by two metrics: gravimetric energy density (kwh/kg) and battery cost (kwh/$). We follow the GML formalism further by estimating the technological improvement rates of these metrics for the simulation analyses by assuming that the technological performance trends follow the exponential

function over time as shown equation (4). Previous studies have empirically confirmed that the exponential relationship between technological performance and time applies in a wide variety of technological domains including energy storage [40-47].

$$P_i(t) = P_i(t_0)\exp(k_i(t - t_0)) \quad (4)$$

Where $P_i(t)$ and $P_i(t_0)$ represent technological performance at time $t$ and at a reference time $t_0$, respectively, for technological domain $i$, the exponential constant $k_i$ denotes the percentage change of performance per year and can be referred to as the technological improvement rate.

Rechargeable batteries previously used in BEVs include lead acid (Pb-acid), Nickel-Cadmium (NiCd), Nickel-Metal-Hydride (NiMH), Lithium-ion (Li-ion), Lithium-ion Polymer, and Sodium Nickel Chloride (NaNiCl), but Li-ion batteries are considered to be the standard for modern BEVs. Li-ion batteries have seen significant increases in energy density and cost reductions since Sony released the first commercially available Li-ion battery in 1991. Compared to other battery technologies, Li-ion batteries have been considered best for BEVs because of a large power storage capacity with excellent specific energy and energy density. However, Li-ion batteries also have some limitations when used for BEVs such as high cost, a potential for overheating, and a limited life cycle [48]. There are some promising alternative candidates for energy storage in BEVs such as metal-air batteries which have high energy density. Even though there has been considerable technological progress in the promising batteries, the practical application is still challenging [49]. Therefore, this study focuses on Li-ion batteries.

Technological performance over time in Li-ion batteries is plotted in Figure 3. The reasonably high value of $R^2$ of the exponential fit indicates that the generalized Moore's Law works adequately for Li-ion batteries as it does for the large number of other domains where studies have been made.

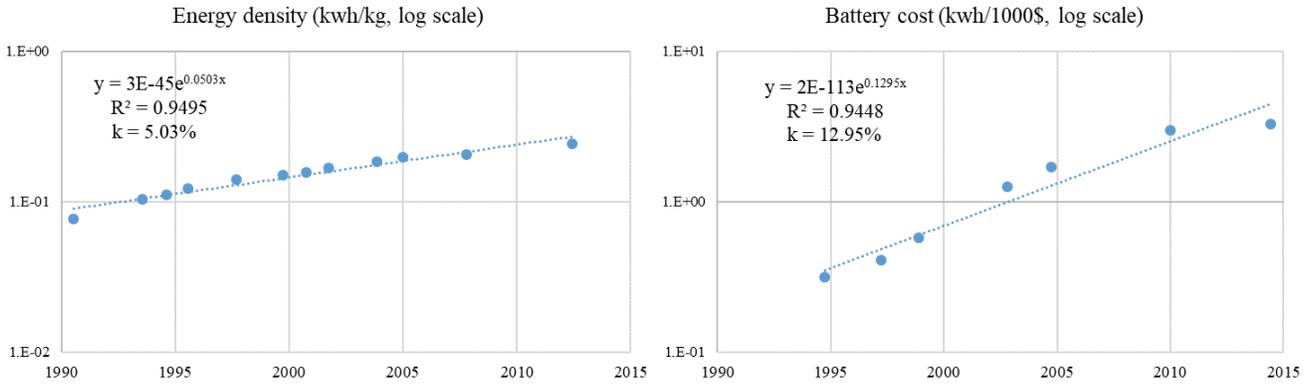

**Figure 3. Technological progress in Li-ion battery technology [50].**

## 3. Results and discussion

### 3.1. Simulation result 1: Trade-offs in ICEVs and BEVs

We simulate how the vehicle attributes and relative value index of hypothetical low performance compact and high performance large BEVs change when their battery sizes are changed from 1 to 300 kWh (holding other input parameters constant) based on NREL's FASTSim software. Then, we also conduct the similar simulation analyses for low performance compact and high performance large ICEVs as the energy storage capacity of ICEV fuel tank is changed from 1 to 800 kWh (about 0-24 gallons of gasoline). The simulation results for the low performance compact BEV and ICEV are shown in Table 4 and Table 5, respectively. The results for high performance large BEV and ICEV are shown in Appendix B.

**Table 4. Simulation result for low performance compact BEV**

| Battery size (kWh) | 1 | 10 | 30 | 50 | 80 | 100 | 200 | 300 |
|---|---|---|---|---|---|---|---|---|
| Range(miles) | 3.61 | 32.62 | 95.49 | 155.39 | 240.16 | 293.61 | 529.09 | 722.55 |
| Energy consumption(kwh/mile) | 0.28 | 0.31 | 0.31 | 0.32 | 0.33 | 0.34 | 0.38 | 0.42 |
| Zero Sixty(s) | 8.11 | 8.29 | 8.71 | 9.13 | 9.71 | 10.15 | 12.26 | 14.35 |
| Interior volume(cubic feet) | 111.64 | 111.13 | 110.00 | 108.87 | 107.17 | 106.04 | 100.39 | 94.74 |
| Depreciation($/year) without tax credit | 5225 | 6035 | 7835 | 9635 | 12335 | 14135 | 23135 | 32135 |
| Depreciation($/year) with tax credit* | 3725 | 4535 | 6335 | 8135 | 10835 | 12635 | 21635 | 30635 |
| fuel costs($/year) | 354 | 392 | 402 | 412 | 426 | 436 | 484 | 531 |
| Cost of time spent refueling($/year) | 101274 | 12664 | 5432 | 4017 | 3259 | 3025 | 2676 | 2690 |
| Relative value index without tax credit | 1.74 | 9.48 | 12.48 | 11.45 | 9.30 | 8.02 | 4.20 | 2.52 |
| Relative value index with tax credit* | 1.76 | 10.29 | 14.02 | 12.82 | 10.26 | 8.76 | 4.46 | 2.63 |

* The federal Internal Revenue Service tax credit is $7,500 per new BEV purchased for use in the U.S.

**Table 5. Simulation result for low performance compact ICEV**

| Fuel storage (kWh) | 1 | 50 | 100 | 200 | 300 | 400 | 500 | 600 | 700 | 800 |
|---|---|---|---|---|---|---|---|---|---|---|
| Fuel tank size (gal) | 0.0 | 1.5 | 3.0 | 5.9 | 8.9 | 11.9 | 14.8 | 17.8 | 20.8 | 23.7 |
| Range(miles) | 0.9 | 46.3 | 92.4 | 184.3 | 275.7 | 366.6 | 456.9 | 546.8 | 636.1 | 724.9 |
| Energy consumption(kwh/mile) | 1.079 | 1.080 | 1.082 | 1.085 | 1.088 | 1.091 | 1.094 | 1.097 | 1.100 | 1.104 |
| Zero Sixty(s) | 9.20 | 9.23 | 9.26 | 9.31 | 9.35 | 9.39 | 9.41 | 9.45 | 9.52 | 9.58 |
| Interior volume(cubic feet) | 111.73 | 111.54 | 111.35 | 110.96 | 110.58 | 110.19 | 109.81 | 109.42 | 109.04 | 108.65 |
| Depreciation($/year) | 3997 | 3998 | 3999 | 4001 | 4003 | 4006 | 4008 | 4010 | 4012 | 4014 |
| fuel cost($/year) | 895 | 896 | 897 | 900 | 902 | 905 | 907 | 910 | 912 | 915 |
| Cost of time spent refueling($/year) | 84995 | 1735 | 886 | 461 | 320 | 249 | 206 | 178 | 158 | 143 |
| Relative value index | 1.82 | 24.61 | 28.09 | 30.00 | 30.57 | 30.70 | 30.76 | 30.65 | 30.43 | 30.19 |

The simulation results for all four types of hypothetical vehicles (low performance compact BEV and ICEV, high performance large BEV and ICEV) are summarized in Figure 4 for direct comparison. Figure 4-(a), 4-(b), and 4-(c) show the relationships of interior volume, acceleration time, and total ownership cost per mile with driving range in four hypotherical vehicles as battery size or fuel tank size is changed. Figure 4-(a) shows that the interior volume decreases more on BEVs than on ICEVs as the driving range increases because batteries takes more volume to store the same amount of energy than gasoline. Similarly, Figure 4-(b) shows that the acceleration performance decreases more on BEVs than on ICEVs as the driving range increases because batteries takes more weight to store the same amount of energy than gasoline. However, when the driving range is low avoiding much battery weight, the acceleration performance of BEVs is better than ICEVs because electric motors work more efficienctly and powerfully than an internal combustion engine.

Figure 4-(c) shows that the total ownership cost per mile increases more on BEVs than on ICEVs as the driving range increases. Total ownership cost per mile in Figure 4-(c) can be decomposed into depreciation per mile, fuel cost per mile, and cost of time spent refueling or recharging per mile. The decomposed ownership cost per mile for the four hypothetical vehicles with driving range of 100, 200, 300, and 400 miles are shown in Figure 5. Figure 5 indicates that BEVs have higher ownership cost compared to ICEVs because BEVs have higher depreciation and cost of time spent recharging than ICEVs. BEVs have lower fuel cost than ICEVs, but it is not enough to offset high depreciation and cost of time spent recharging. Unlike ICEVs, total ownership cost of BEVs increases rapidly as driving range increases more than offsetting the decreases in the cost of time spent recharging because BEV batteries are much more expensive than ICEV fuel tank expansion.

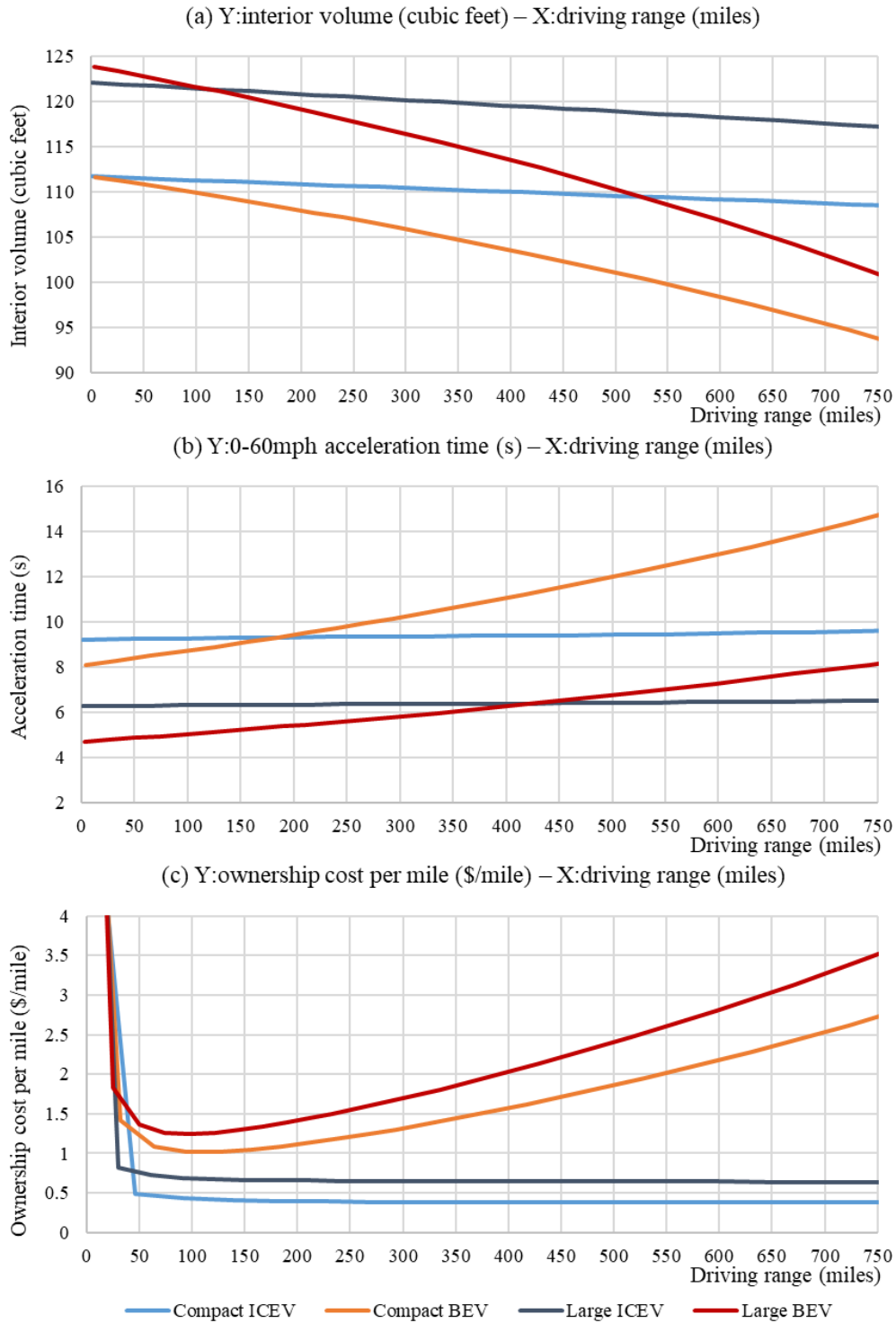

**Figure 4. Core attributes as a function of driving range (miles)**

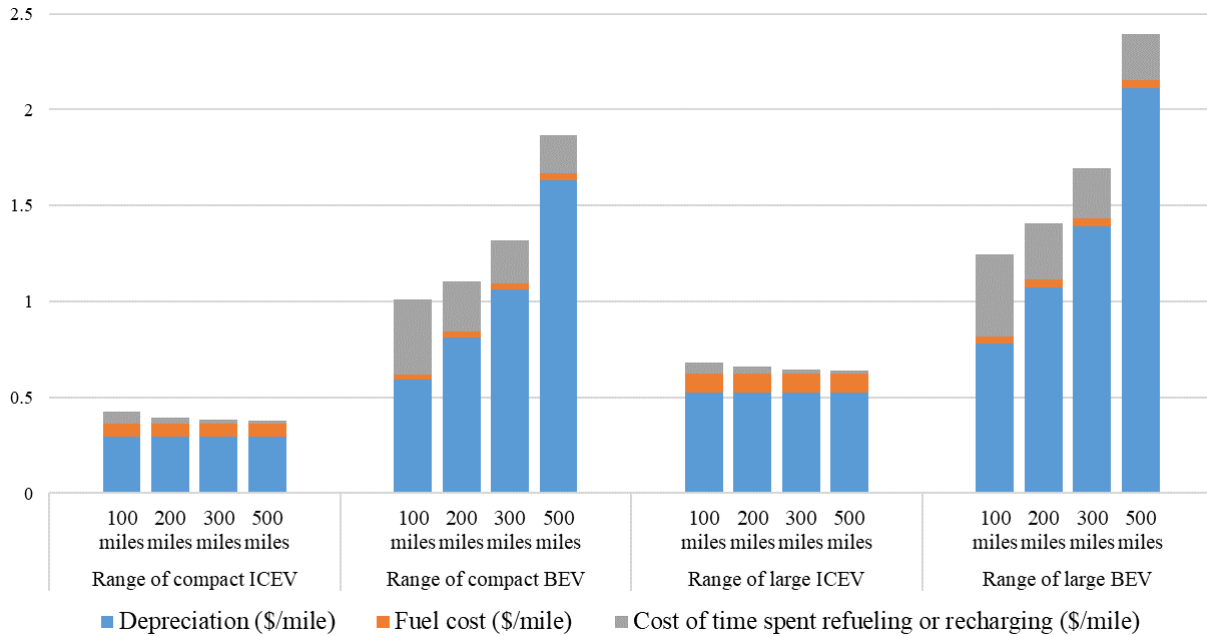

**Figure 5. Decomposition of ownership cost per mile**

Overall comparisons of both sets of BEV and ICEV vehicles are given in Figures 6 and 7. Figure 6 presents the relationship between relative value index and driving range in low performance compact BEV and ICEV, high performance large BEV and ICEV over the full set of driving ranges; the figure shows that the effect of the tax credit $7,500 on the relative value of BEVs is not very significant. This figure also indicates that the optimal driving range to maximize relative value index of BEVs is in the range of 70-100 miles consistent with actual offerings to consumers. The normalized interior volume, acceleration performance (1/acceleration time), cost advantage (1/ownership cost), and relative value index of ICEVs to those of BEVs are shown in Figure 7. This figure demonstrates that the increasing relative advantage of ICEVs with driving range seen in Figure 6 is mostly due to the growing cost advantage as range is increased.

Overall, Figure 6 and Figure 7 indicate that the relative value of BEVs is lower than that of ICEVs because BEVs have high battery cost and high cost of time spent recharging despite high torque, high fuel efficiency and low fuel cost. Figure 6 and figure 7 also show that the relative value differences between BEVs and ICEVs becomes larger due to the high battery cost (the influence of changes in interior volume and accerleration performance is not as significant) if battery size of BEVs is increased and BEVs' driving range become similar to ICEVs on the market. The relative value differences between BEVs and ICEVs are seen in Figure 6 to be less in high performance large cars than in low performance compact cars because BEVs can achieve high accelaration performance more easily than ICEVs.

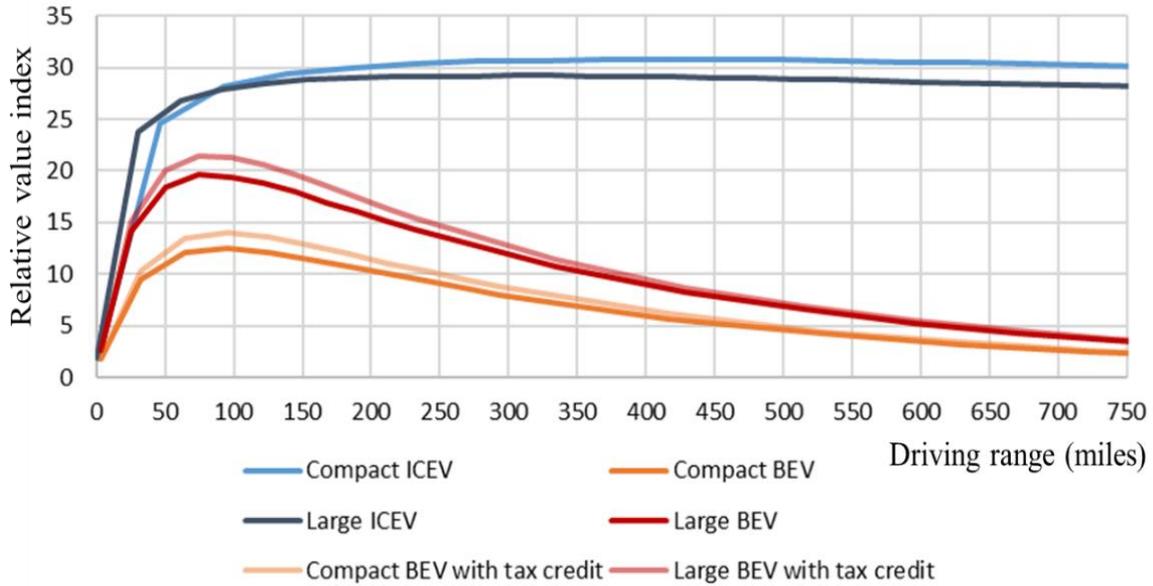

**Figure 6. Relative value index as a function of driving range**

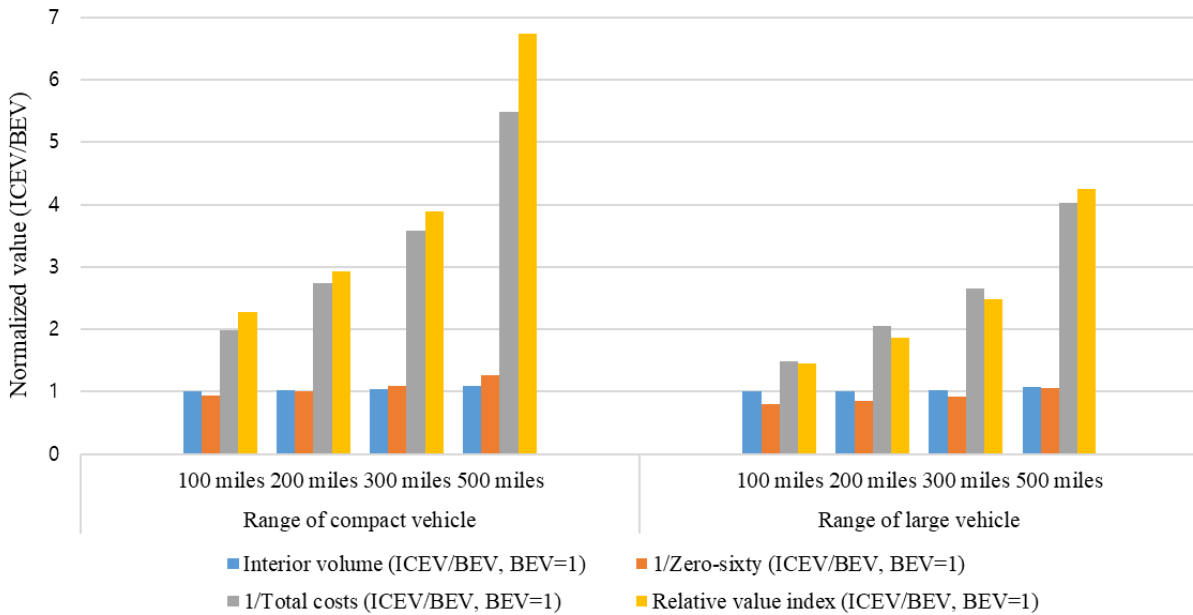

Note: Interior volume, acceleration performance (1/acceleration time), cost advantage (1/ownership cost), and relative value index of ICEVs normalized to those of BEVs

**Figure 7. Normalized attributes and relative value index**

## 3.2. Simulation result 2: technological progress in battery technology

As an extension of the preceding analysis, we analyze when the relative value index of BEVs might exceed ICEVs as the energy density and cost of Li-ion batteries improve according to ongoing technological progress. Specifically, energy density and cost of Li-ion battery technology have been improved about 5% per year and 13% per year, respectively (Figure 3). The energy density of Li-ion batteries has steadily increased mostly because of new materials for the cathode, anode, and electrolyte [14, 50]. The cost of Li-ion batteries also has decreased steadily and rapidly but the technological improvement rate for cost could be overestimated by the recent increase in production to meet demand for mobile consumer devices. Such production scale changes decrease cost and temporarily increase the technological improvement rate ($k$) for cost but do not effect $k$ for metrics such as energy density. Thus, we assume that Li-ion battery energy density (kwh/kg) will increase by 5% per year and Li-ion battery cost (kwh/$) will decrease by 10% per year and then simulate the vehicle attributes and relative value index of the hypothetical BEVs and ICEVs when their battery size and fuel tank size is changed in 2025, 2035, and 2050 (holding other input parameters constant).

Figure 8, 9, and 10 present the forecast simulation results. Specifically, Figure 8 shows the relationship between relative value index and driving range in low performance compact BEV and ICEV, high performance large BEV and ICEV. Normalized Interior volume, acceleration performance (1/acceleration time), cost advantage (1/ownership cost), and relative value index of ICEVs to those of BEVs with 300 mile driving range in 2016, 2025, 2035, and 2050 are shown in Figure 9. In addition, the decomposed ownership cost per mile for the BEVs and ICEVs with driving range of 300 miles in 2016, 2025, 2035, and 2050 are shown in Figure 10.

Figure 8–(a) and 8-(b) indicate that by 2050, high performance large BEVs could have higher relative value index than high performance large ICEVs because of technological improvements in Li-ion batteries, but low performance compact BEVs will still have significantly lower value than comparable ICEVs. Moreover, these figures also show that the optimal driving range to maximize relative value index of BEVs increases. In 2035 and 2050, the optimal driving range is about 200 miles and 350-400 miles, respectively. Figure 9 and 10 show the source of these forecast improvements-the price of BEVs is getting more competitive with ICEVs because of improvements in the battery cost and the acceleration performance of BEVs is getting better because of improvements in the battery energy density.

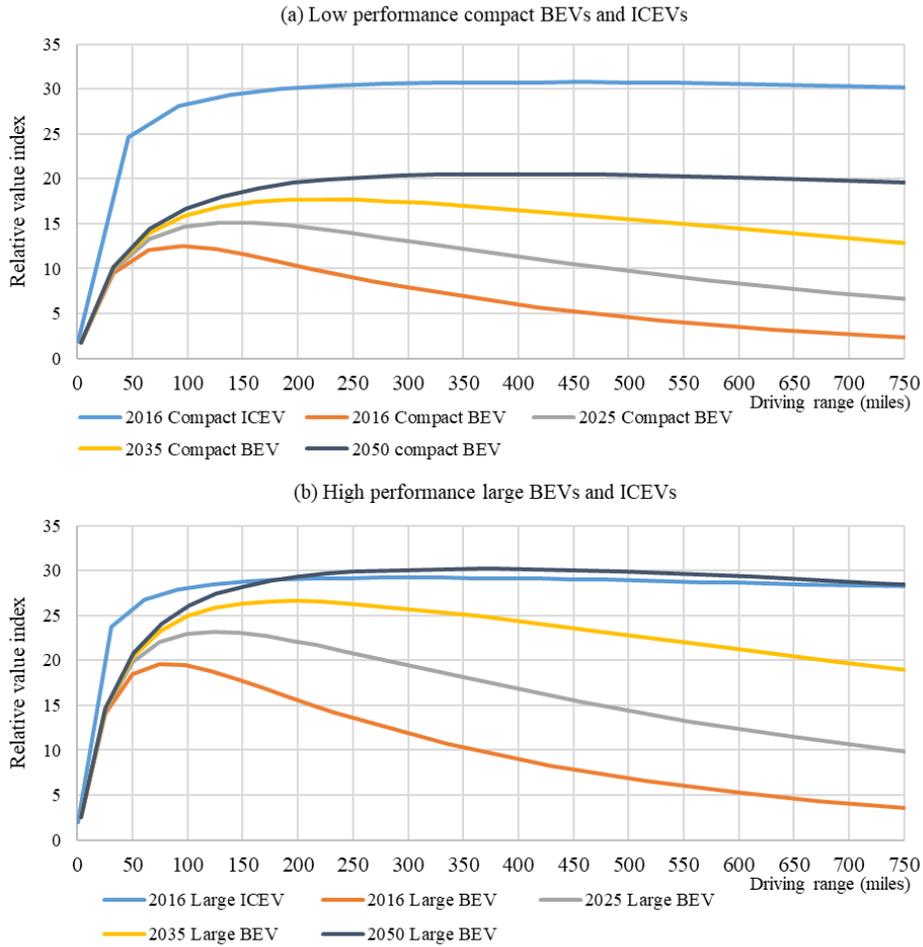

**Figure 8. Relative value index against driving range (miles): 2016, 2025, 2035, 2050.**

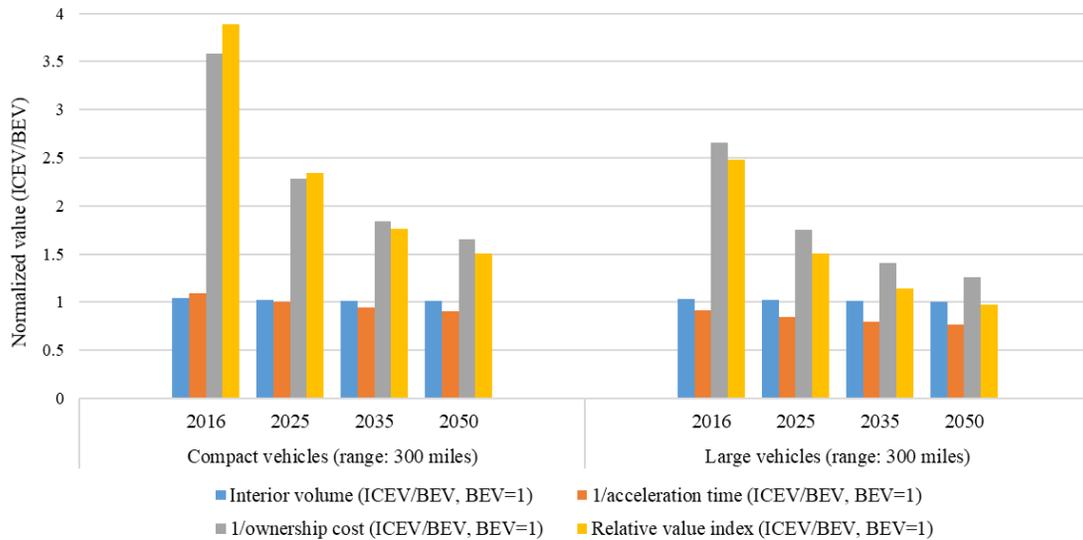

Note: Interior volume, acceleration performance (1/acceleration time), cost advantage (1/ownership cost), and relative value index of ICEVs normalized to those of BEVs.

**Figure 9. Normalized attributes and relative value index**

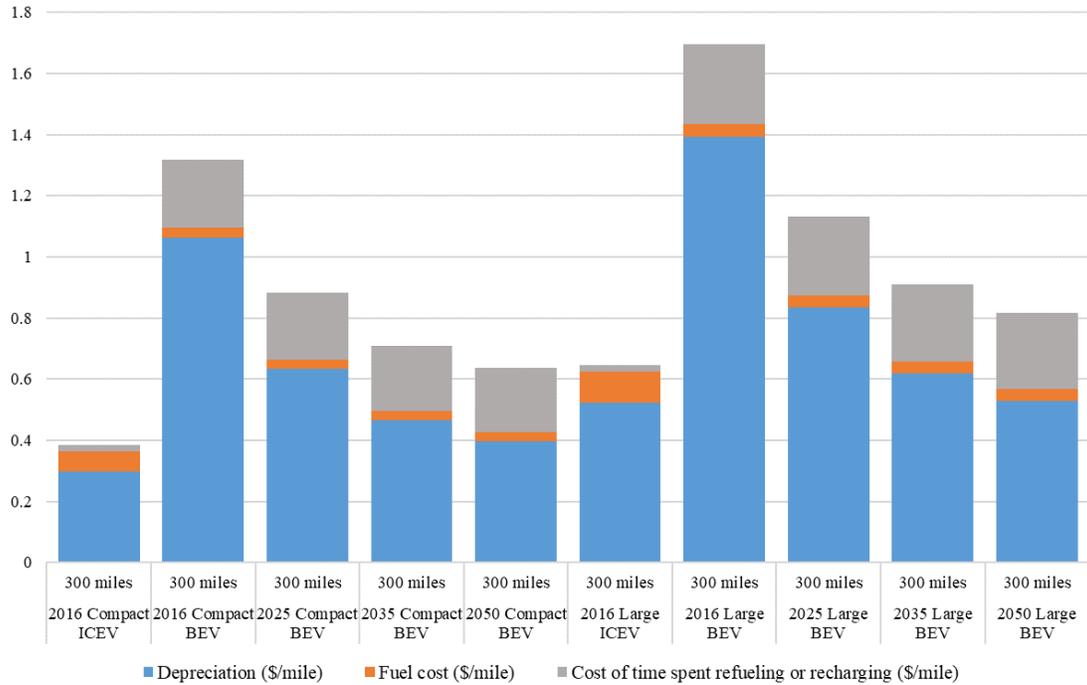

**Figure 10. Decomposition of total ownership cost per mile at 300 mile driving range**

### 3.3. Simulation result 3: sensitivity test

In the above simulations, some key input parameters are fixed at what we judged to be best estimates. However, there is uncertainty about whether reasonable changes in these input paramters might significantly change the implications of our simulation results. Therefore, in this section, we analyze the sensitivity of our simulation results to variation in input parameters (electricity price, gasoline price, percentage of home charging, value of time, charging time, markup factor for battery cost, technological improvement rates for Li-ion battery) to probe this uncertainty. Specifically, we analyze how the relative value index of BEVs and ICEVS change in 2016 and 2050 when these input parameters change from the bulk of the default settings by about ±15%. However, there is good reason [42, 51, 52] to assume that ±50% is a more appropriate range to test for the technological improvement rate ($k$). The sensitivity results are presented in Figure 11 and 12.

From the sensitivity test results for low performance compact BEV and ICEV in Figure 11, we find the overall results are similar to the baseline or default result: the relative user value index does not exceed that of comparable low performance compact ICEV until after 2050 even though the input parameters for sensitivity tests change by ±15% and for the technological improvement rates by ±50%.

The change of input parameters like electricity price, gasoline price, value of time, charging time, and markup factor for battery cost had only minor effects on the relative value index of low performance compact BEV and ICEV. On the other hand, the change of home or workplace charging percentage affected the relative value index of low performance compact BEV significantly. That is, BEVs can give much greater value to consumers who charge their BEVs mostly at home or at work. While this is might be considered evidence of possible lack of robustness in our simulation, we note an important mitigating factor. Note first that the sensitivity to this parameter is only large at low driving range because at longer ranges, the high battery costs dominate in keeping the BEV relative value index low. Since low range BEVs are probably only suited for home to work travel and only for some people, this sensitivity to home/work charging does not seem to suggest very large scale substitution which is what we are trying to assess in this paper. Comparing Figures 11-(a) and 11-(b), we again see that technological progress does act to narrow the relative value gap between ICEVs and BEVs so sensitivity to variation of the technological improvement rate by ±50% is expected and is seen in Figure 11-(b). However, even this test supports the conclusion that mass market substitution is not likely for Li-ion BEVs even in 2050. Overall, the sensitivity results indicate that the relative user value of low performance compact BEV does not suggest it overtaking comparable low performance compact ICEV by 2050 even if drivers charge the BEV more than 90% at home or at work and even if the technological improvement rate increases 50% over its trend for the past 20 years. This is further suggestive that the mass of the automotive market is unlikely to transition to Li-ion BEVs by 2050.

      The sensitivity test results for high performance large vehicles in Figure 12 also show that the changes in electricity price, gasoline price, value of time, charging time, and markup factor for battery cost did not change the relative user value index of high performance large BEV and ICEV significantly. However, the percentage of home or workplace charging and the uncertainty in technological improvement again have significant impacts on the relative user value index of high performance large BEV. Specifically, in 2016, if drivers charge their high performance large BEVs more than 90% at home or at work, it is possible that the relative user value index of the BEVs is similar to or higher than that of comparable high performance large ICEVs when the driving range of BEVs is very short. Moreover, in 2050, the relative user value of high performance large BEVs could be much higher or lower than that of comparable high performance large ICEVs depending on the drivers' home or workplace charging percentage and depending upon the realized technological improvement rate between now and 2050. Thus, drivers' home or work place charging percentage and the upcoming technological improvement rate for

Li-ion batteries are significant uncertainties to the long term value comparison in high performance vehicles. Overall, the results in Figure 12 indicate that is is highly likely that Li-ion BEVs will be competitive with or superior to ICEVs in the high performance luxury automotive market. However, these results are not as environmentally important as those in Figure 11 because an expanded luxury market will not serve the environmental goals desired from BEVs.

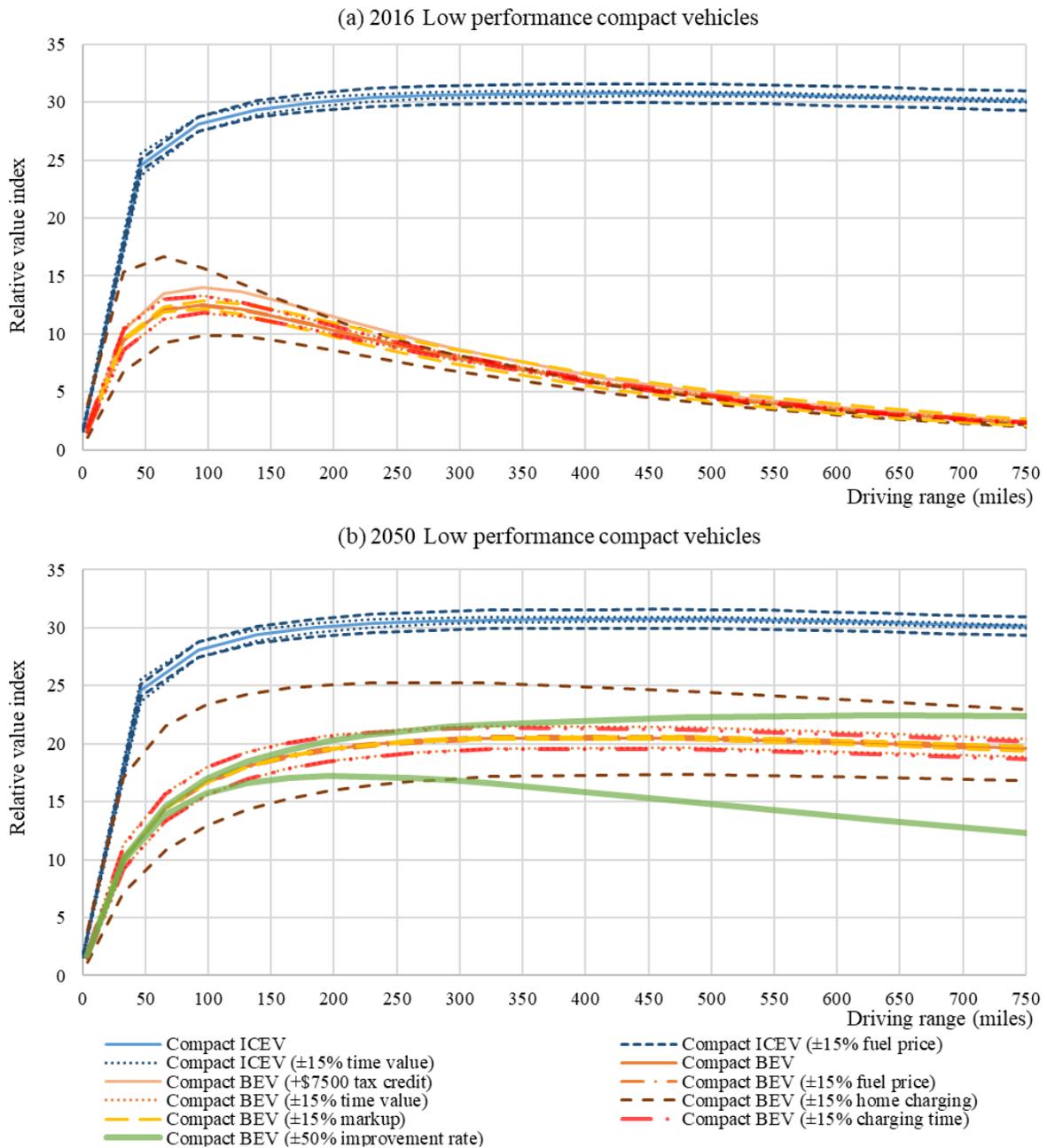

**Figure 11. Sensitivity test results for low performance compact vehicles in 2016 and 2050**

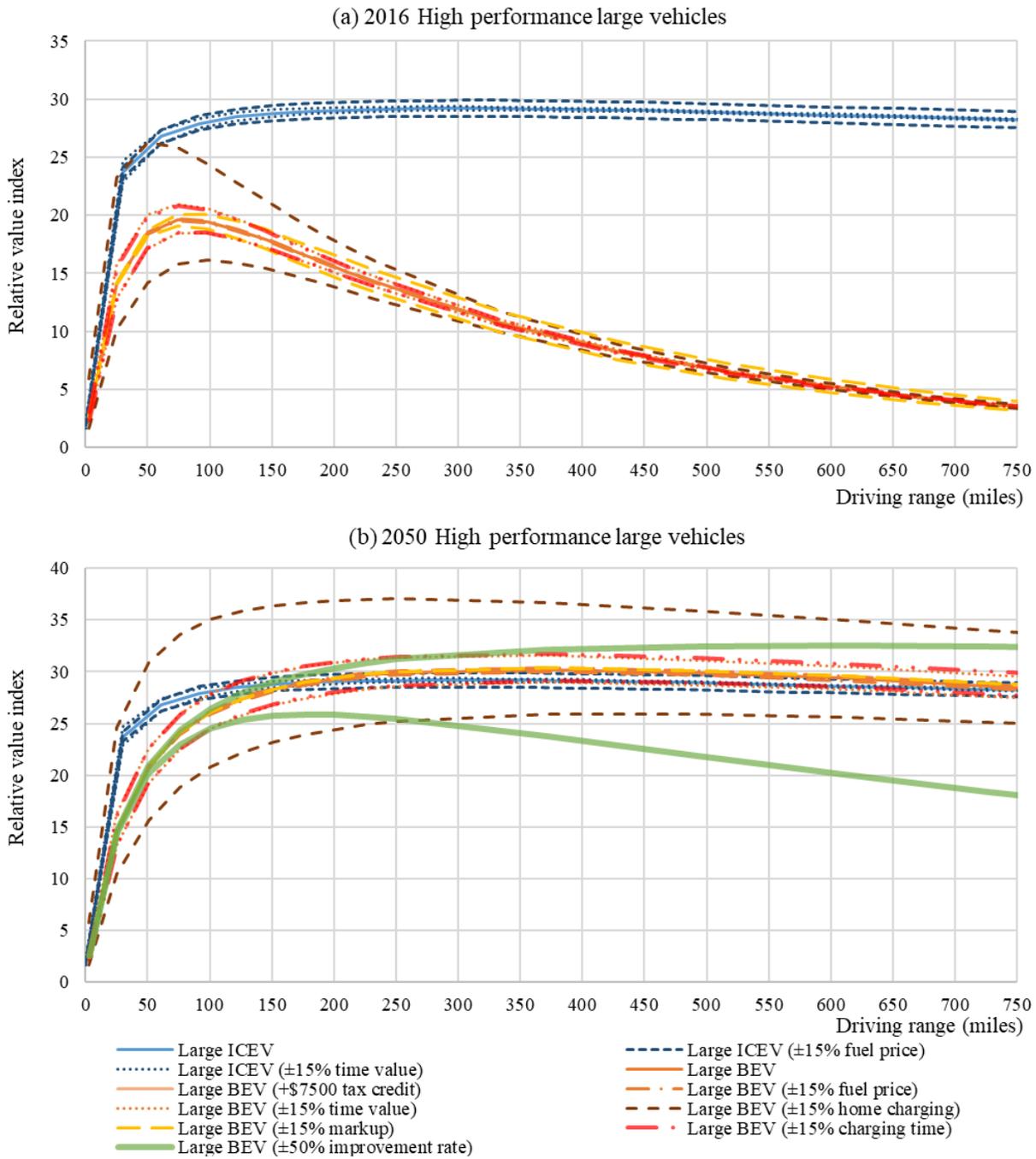

**Figure 12. Sensitivity test results for high performance large vehicles in 2016 and 2050**

**4. Discussion and conclusions**

This study assessed and forecasted whether and when design changes and technological improvements related to the major challenges in driving range and battery cost will make the user value of BEVs greater than the user value of ICEVs. In the initial part of study, we simulated changes in the vehicle attributes and relative value index of hypothetical low performance compact and high performance large BEVs and ICEVs according to changes in their battery size and fuel tank size (holding other input parameters constant at the 2016 level).

The simulation and analysis indicates that the relative value index of BEVs is lower than that of ICEVs because BEVs have high battery cost and high cost of time spent recharging despite high torque, high energy efficiency, and low fuel cost. Thus, in the current situation, U.S. average consumers are less likely to find that BEVs are more valuable than ICEVs. In addition, although the U.S. government's tax credit for BEVs increases the relative value of BEVs compared to ICEVs, it does not seem to be enough to make the relative value of BEVs similar to or larger than ICEVs particularly in the mainstream non-luxury segment (represented in this study by the low performance compact vehices). These results are consistent with the fact that BEVs are not yet widely accepted in the automobile market, unlike ICEVs. These results are also consistent with the relative success thus far of higher performance large luxury BEVs- specifically Tesla.

The simulations also showed that unlike ICEVs, the relative value index of BEVs decreases sharply as driving range increases because the marginal cost of a BEV battery is much more expensive than expansion of an ICEV fuel tank (the influence of changes in interior volume, accerleration performance, and cost of time spent recharging are much less significant). The relative value differences between BEVs and ICEVs becomes larger if battery size of BEVs is increased and BEVs' driving range become similar to ICEVs on the market. Accordingly, the optimal driving range to maximize the relative value index of BEVs was observed conspicuously in the range of 70 – 100 miles. That is, the simulations suggest that the driving range of BEVs should be designed to be about 70 – 100 miles to maximize the value of BEVs offered to U.S. average consumers. The fact that the driving range of most BEVs on the market today is consistent with these results suggests support for an assertion that the simulation results and relative value index used in this study are reasonable approximations of reality.

Furthermore, our simulation results showed that the relative value differences between BEVs and ICEVs is less in high performance large cars than in low performance compact cars because BEVs can achieve high accerlaration performance more easily than ICEVs. This is consistent with and possibly explains why Tesla's high performance large BEVs, despite their very high price, sell better than other low performance compact BEVs in the U.S. automobile market. The simulation suggests that it is structurally advantageous to sell high performance luxury cars rather than low performance non-luxury cars in the BEV market at this time.

In the forecasting part of this study, we simulated the changes in vehicle attributes and relative value index of the hypothetical BEVs as the energy density and cost of Li-ion batteries improve because of ongoing technological change. The ownership cost of BEVs becomes cheaper because of improvements in the battery cost and the acceleration performance of BEVs improves because of improvements in the battery energy density. From the simulation analysis, we found that by 2050, high performance large BEVs are likely to have higher relative value than high performance large ICEVs because of technological improvements in Li-ion batteries. However, in the mass market represented in this study by low performance compact BEVs, Li-ion BEVs would still have significantly lower value than comparable ICEVs even after the same (and possibly higher) level of technological improvements. This suggests that expectations of near-total substitution of BEVs for ICEVs by 2050 (or earlier) is potentially misplaced and wrong. This conclusion is the major finding of the present study and since it is a forecast the limitations must be especially carefully considered.

There are a wide variety of limitations on the current study which can be analyzed. Table 6 gives the implications of four major sources of limitations which we now discuss.

**Table 6. Limitations for the current study and Implications to conclusions**

| Limitation | Possible Implications |
|---|---|
| 1. Relative value index and simulation of trade-offs | If incorrect, the forecast as well as the current situation description could be wrong |
| 2. Global vs U.S. market considerations | If the global demand more strongly favors BEVs than does the U.S., the environmental impact of the forecast is incorrect |
| 3. Other vehicle technological change | If batteries other than Li-ion or other vehicle technologies evolve before 2050, the negative implications of the forecast may be misleading |
| 4. Future automotive demand structure | If car sharing and/or autonomous vehicles become prevalent by 2050, the forecast might be wrong |

The first limitation to the forecast is that it is made using a model for relative value index and that this model is analyzed by a simulation model for BEVs and ICEVs. If either of these models does not adequately represent reality, the forecast is highly suspect. The simplicity of the relative value index (minimum attributes for describing the essential attributes of BEVs sand ICEVs) and the relative simplicity of the simulation compared to the complex reality of a vehicle denotes that this limitation is potentially significant. While we cannot simply assert that this limitation is not potentially important, we feel that the consistency of results with current market and design choices suggest that their description of reality is acceptable. In particular, our results suggest that high performance luxury BEVs are more competitive with comparable ICEVs than are mass-market cars which is aligned well with market reality in the U.S. automotive market. Similarly, our results and reality are also aligned with the indication that the highest value vehicles that can be offered in the mass market have a driving range of 70-100 miles. Additionally, the choice of specific values of parameters –particularly the rate of technological improvement- does not greatly change the forecast and thus also does not change our primary conclusion.

The second listed limitation in Table 6 is that our study focuses on the U.S. market results and the forecast for global automotive demand is much more significant from an environmental standpoint. There are two arguments that suggest this limitation is not likely to upset the conculsions in the paper. First, the U.S. market for BEVs is currently the most developed one and secondly there are no known reasons for consumers elsewhere to accept attributes that differ between BEVs and ICEVs differently from U.S. users.

The third source of limitation to the current study is potentially broader than often recognized. The prior support for the generalized Moore's Law, the data for Li-ion batteries and the fact that a 50% positive deviation in the rate of improvement does not overturn the mass market conclusion are strong evidence that Li-ion BEVs are not likely to overtake ICEVs in the mass automotive market by 2050. However, other technologies may achieve the same environmental goal and they are not included in our analysis. In particular, other battery systems than Li-ion are certainly feasible before 2050. Although it is possible that such batteries have higher the technological improvement rates ($k$) than Li-ion batteries, significantly more rapid technological improvement is not likely for such batteries based upon the long-term performance data on a variety of battery systems [44]. However, capacitors are an electrical energy storage technology that is known to have a significantly higher $k$ [44] and indeed capacitors have been forecast to surpass batteries in energy storage capability by the 2030s [41]. In addition, fuel cells are improving more rapidly than batteries [40] so it is possible that $H_2$ fuel cell vehicles could be superior to Li-ion BEVs before 2050 (however $H_2$ storage is also challenging). Thus, our focus on Li-ion battery technology is a limitation that probably does not overturn our basic conclusion for BEVs in general; however it is likely to be a serious limitation if one considers other alternatives. In particular, there appears to be a decent chance that Capacitor Electric Vehicles (CEVs) will be superior to ICEVs before 2050. Of course, predicting this with any certainty with zero CEVs sold as of now would be highly problematic.

The fourth source of potential limitation listed in Table 6 concerns the fact that our analysis to some extent assumes that the automotive market structure for 2050 is similar to today's relative to attribute importance. In general, this is not necessarily correct and it is possible over the next 30 years that evolution of transportation will occur that markedly changes the structure and desirability of attributes. For example, extremely high carbon taxes would enable development of carbon-free electricity and possibly be able to offset the value shortfall of BEVs relative to ICEVs. Although this appears unlikely based upon the past several decades, political changes in will cannot be ruled out. Beyond the political uncertainty, there are also some other technological changes –in particular driverless vehicles and shared vehicles- that appear likely to grow in importance before 2050. Neither of these changes appear to favor the attributes naturally superior/inferior in BEVs; indeed, higher performance and low driving range may well be less favored with shared and/or driverless vehicles making BEVs even less competitive in such environments.

The limitations of any forecast more than 30 years into the future cannot be entirely eliminated and that is true for the forecast in this study. Our preceding analysis indicates that the greatest uncertainty relative to to Li-ion BEV competitiveness up to 2050 involves large political actions that are possible but are not foreseen at the present time. The other major uncertainty this analysis uncovers is that other technologies –in particular CEVs- may well achieve the desired move from ICEVs even before 2050. This positive conclusion leads one to stress the desirability of not locking into a desirable alternative too early since we do not want to make a good approach prevent a better one from evolving.

**Acknowledgments**

The authors gratefully acknowledge the support of the SUTD/MIT International Design Center.

**Appendix A. Simulation input parameters**

**Table A.1. Simulation input parameters for low performance compact and high performance large ICEVs**

| Classification | Parameter | low performance compact ICEV | high performance large ICEV |
|---|---|---:|---:|
| Vehicle | Drag coefficient | 0.29 | 0.28 |
|  | Frontal area (m$^2$) | 2.07 | 2.14 |
|  | Vehicle glider mass (kg) | 768.77 | 809 |
|  | Vehicle center of gravity height (m) | 0.53 | 0.53 |
|  | Wheel base (m) | 2.59 | 2.82 |
| Engine | Engine power (kW) | 100 | 250 |
| Fuel tank | **Variable: Fuel storage energy (kWh)** | **From 1 to 800** | **From 1 to 800** |
| Wheel | Number of wheels | 4 | 4 |
|  | Tire radius (m) | 0.317 | 0.334 |

# Appendix B. Simulation results

**Table B.1. Simulation result for high performance large BEV**

| Battery size (kWh) | 1 | 10 | 30 | 50 | 80 | 100 | 200 | 300 |
|---|---|---|---|---|---|---|---|---|
| Range(miles) | 2.85 | 25.23 | 74.29 | 121.58 | 189.38 | 232.69 | 428.92 | 596.28 |
| Energy consumption(kwh/mile) | 0.35 | 0.40 | 0.40 | 0.41 | 0.42 | 0.43 | 0.47 | 0.50 |
| Zero Sixty(s) | 4.71 | 4.78 | 4.95 | 5.12 | 5.37 | 5.55 | 6.40 | 7.24 |
| Interior volume(cubic feet) | 123.90 | 123.39 | 122.26 | 121.13 | 119.43 | 118.30 | 112.65 | 107.00 |
| Depreciation($/year) without tax credit | 6934 | 7744 | 9544 | 11344 | 14044 | 15844 | 24844 | 33844 |
| Depreciation($/year) with tax credit* | 5434 | 6244 | 8044 | 9844 | 12544 | 14344 | 23344 | 32344 |
| fuel cost($/year) | 448 | 507 | 517 | 526 | 541 | 550 | 597 | 644 |
| Cost of time spent refueling($/year) | 128141 | 16378 | 6982 | 5135 | 4132 | 3817 | 3301 | 3260 |
| Relative value index without tax credit | 2.62 | 14.14 | 19.58 | 18.77 | 16.04 | 14.25 | 8.26 | 5.28 |
| Relative value index with tax credit* | 2.65 | 15.05 | 21.47 | 20.58 | 17.44 | 15.39 | 8.72 | 5.50 |

*The federal Internal Revenue Service tax credit is $7,500 per new BEV purchased for use in the U.S.

**Table B.2. Simulation result for high performance large ICEV**

| Fuel storage size (kWh) | 1 | 50 | 100 | 200 | 300 | 400 | 500 | 600 | 700 | 800 |
|---|---|---|---|---|---|---|---|---|---|---|
| Fuel tank size (gal) | 0.03 | 1.48 | 2.97 | 5.93 | 8.90 | 11.87 | 14.84 | 17.80 | 20.77 | 23.74 |
| Range(miles) | 0.61 | 30.53 | 61.00 | 121.76 | 182.28 | 242.56 | 302.61 | 362.42 | 421.99 | 481.33 |
| Energy consumption(kwh/mile) | 1.636 | 1.638 | 1.639 | 1.643 | 1.646 | 1.649 | 1.652 | 1.656 | 1.659 | 1.662 |
| Zero Sixty(s) | 6.271 | 6.283 | 6.293 | 6.314 | 6.331 | 6.347 | 6.360 | 6.375 | 6.390 | 6.409 |
| Interior volume(cubic feet) | 122.12 | 121.93 | 121.74 | 121.35 | 120.96 | 120.58 | 120.19 | 119.81 | 119.42 | 119.04 |
| Depreciation($/year) | 7050 | 7051 | 7052 | 7054 | 7056 | 7058 | 7060 | 7062 | 7064 | 7066 |
| fuel cost($/year) | 1357 | 1358 | 1359 | 1362 | 1365 | 1367 | 1370 | 1373 | 1375 | 1378 |
| Cost of time spent refueling($/year) | 128892 | 2630 | 1342 | 698 | 483 | 376 | 312 | 269 | 238 | 215 |
| Relative value index | 1.91 | 23.73 | 26.78 | 28.47 | 28.97 | 29.14 | 29.19 | 29.15 | 29.07 | 28.96 |